\documentclass[useAMS,usenatbib,usegraphicx]{mn2e}

\title[The discovery of two new galaxy--galaxy lenses from the
SDSS]{The discovery of two new galaxy--galaxy lenses from the
SDSS\thanks{Based upon observations obtained using the Magellan
telescopes at the Las Campanas Observatory and the European Southern
Observatory (programme IDs: 73.A-0503 and 74.A-0481), Chile.}}
\author[J. P. Willis, P. C. Hewett and
S. J. Warren]{J. P. Willis$^{1}$\thanks{E-mail: jwillis@uvic.ca},
P. C.  Hewett$^{2}$ and S. J. Warren$^{3}$\\ $^{1}$Department of
Physics and Astronomy, University of Victoria, Elliot Building, 3800
Finnerty Road, Victoria, V8V 1A1, BC, Canada\\ $^{2}$Institute of
Astronomy, Madingley Road, Cambridge, CB3 OHA, UK\\ $^{3}$Blackett
Laboratory, Imperial College of Science Technology and Medicine,
Prince Consort Road, London SW7 2BW, UK}

\begin{document}

\date{Accepted. Received;}

\pagerange{\pageref{firstpage}--\pageref{lastpage}} \pubyear{2005}

\maketitle

\label{firstpage}

\begin{abstract}

The gravitational lens configuration where a background galaxy is
closely aligned with a foreground galaxy, can provide accurate
measurement of the dark matter density profile in the foreground
galaxy, free of dynamical assumptions.  Currently only three such
galaxy--galaxy lenses are known where the lensed source has a
confirmed redshift and is reasonably bright at optical wavelengths,
and therefore suitable for observations with the HST Advanced Camera
for Surveys (ACS).  Two of these were discovered by noting an
anomalous emission line (from the source) in the spectrum of a massive
early--type galaxy (the lens).  To find further galaxy--galaxy lenses
suitable for ACS imaging we have looked for anomalous emission lines
in the luminous red galaxy (LRG) subsample of the SDSS DR3
spectroscopic database.  Our search methodology has similarities to
that applied by Bolton et al.  (which has had recent success), but
extends the upper redshift limit for lensed sources to $z\simeq 1.4$.
Here we report follow-up imaging and spectroscopic observations of two
candidates, confirmed as gravitational lenses by the detection of
multiple images in the line of [OII] $\lambda\lambda$3726,3729.  In
the first system, J145957.1-005522.8, the lens at $z=0.58$, consists
of two LRGs.  The anomalous emission line is confirmed as [OII] by the
detection of the corresponding H$\gamma$ line, providing a source
redshift of $z=0.94$.  In the second system, J230946.3-003912.9, the
lens is a single LRG at $z=0.29$, and the source redshift is $z=1.00$,
confirmed by partially resolving the [OII] doublet.

\end{abstract}


\section{Introduction}

The analysis of accurate rotation curves of low surface brightness
galaxies (e.g. de Blok \& Bosma, 2002) has sparked a debate on whether
the data are consistent with the mass profiles predicted within the
$\Lambda$CDM cosmological paradigm.  The matter is currently
unresolved, and the arguments are centred on the accuracy of the
$\Lambda$CDM predictions at small radii (Navarro et al. 2004), and the
influence of triaxiality on the gas dynamics (Hayashi et al. 2004).
As a means of measuring masses of galaxies, strong gravitational
lensing, where the source is multiply imaged, is free of the
ambiguities of dynamical tracer methods, for example whether the
system is axisymmetric, or in dynamical equilibrium.  In the majority
of cases of strong lensing by galaxies the source is a quasar, and
unresolved.  In such cases, lensing provides a very accurate measure
of the total mass within the Einstein radius, but there are too few
constraints to determine the lens mass profile.  Rusin, Kochanek \&
Keeton (2003) analysed 22 multiply-imaged quasars, and showed how it
is possible to obtain information on the mass profiles of galaxies
from a statistical analysis, by postulating a relation between the
mass and light distributions in the galaxies.  Another successful
approach, developed by Treu \& Koopmans (2004, and references
therein), is to combine lensing information (image positions) and
kinematic information (the velocity dispersion radial profile).  To
some extent this overcomes the individual weaknesses of lensing
studies (only a total mass is measured) and dynamical studies (too
many assumptions), for measuring galaxy mass profiles.

Only a few cases of multiple imaging of galaxies by galaxies are
known, but they offer the prospect of accurate measurement of galaxy
mass profiles free of dynamical assumptions.  For quasars, the image
fluxes do not provide useful constraints on mass models, because of
the possibility they are brightened by microlensing.  But for a
resolved source, viewed through a dust-free galaxy, every resolution
element in the image provides both a flux and a positional constraint.
Therefore the best constraints come from high-resolution images of
Einstein rings, since this maximises the number of resolution
elements.  Optimal inversion of the image requires a method which
solves simultaneously for the source light profile and the lens mass
profile (e.g.  Wallington, Kochanek \& Narayan, 1996, Warren \& Dye,
2003).  Dye \& Warren (2005) analysed the HST image of the Einstein
ring $0047-2808$.  They used a two-component mass model, comprising a
dark-matter halo, and a baryonic component following the light, and
thereby measured the inner slope of the dark matter mass profile
$\gamma=0.87^{+0.69}_{-0.61}$ ($95\%$ confidence).  They obtained
substantially stronger limits than the lensing and dynamics analysis
of the same system by Koopmans \& Treu (2003), which used only the
positions of the four bright peaks in the image as lensing
constraints.  They also obtained constraints on the dark matter
fraction within the Einstein radius from this single system, as good
as the constraints obtained by Rusin et al. (2003) from their analysis
of 22 multiply-imaged quasars, with fewer assumptions. This highlights
the usefulness of galaxy--galaxy lenses.  The analysis also provided
the first measurement, free of dynamical assumptions, of the baryonic
mass-to-light ratio (M/L) in an individual galaxy.

Einstein rings were discovered at radio wavelengths (Hewitt et
al. 1987), and a handful of other radio rings are now known.  In this
paper we are concerned with finding additional suitable galaxy-galaxy
lens targets at optical wavelengths for imaging with the HST Advanced
Camera for Surveys (ACS), in order to measure accurate mass profiles.
The three best lenses for this type of work so far published, where
the ring is substantially complete and reasonably bright at optical
wavelengths, are $0047-2808$ (Warren et al. 1996, 1998), SDSS
J$140228.22+632133.3$ (Bolton et al. 2005), and FOR J$0332-3557$
(Cabanac et al., 2005).  The first two have been observed by HST,
although the images are of relatively low signal-to-noise ratio (S/N).
It would be feasible to obtain high S/N images of all three targets
with ACS, with just a few orbits, and substantially improve on the results
of Dye \& Warren.  We have excluded here rings where the source is an
AGN (the best example is RXS J$1131-1231$, Sluse et al. 2003), because
again the fluxes in the bright images of the nucleus may be affected
by microlensing, and contaminate large sections of the ring.  A number
of other galaxy-galaxy lenses are known, which are fainter and less
extended than the best three, and therefore less useful (e.g.
Ratnatunga et al. 1999, Crampton et al. 2002, Fassnacht et al. 2004).

The lens $0047-2808$ was discovered serendipitously, by noting an
anomalous emission line in the spectrum of an early-type galaxy at
$z=0.485$.  The emission line proved to be Ly$\alpha$ from a
background galaxy at $z=3.6$, shown to be lensed into a partial ring
in a narrow-band image centred on the line.  A targetted search of the
spectra of massive galaxies in large galaxy redshift surveys could
prove an efficient method for discovering further examples (Hewett et
al.  2000).  The spectroscopic catalogue of the Sloan Digital Sky
Survey (SDSS, York et al.  2000) is the best current dataset for this
task by virtue of its size, depth, wavelength coverage, and
resolution, and especially because the relatively large fibre angular
diameter of $3\arcsec$ is larger than the Einstein ring for all but
the most massive galaxies.\footnote{For example, for a singular
isothermal sphere at $z=0.4$, of one--dimensional velocity dispersion
$\sigma_v=220$~km\,s$^{-1}$, the angular diameter of the Einstein ring
is 1\farcs5 and 2\farcs2 for source redshifts of $z=1$ and $z=4$
respectively.}  This ensures that in all cases where the source is
multiply imaged, nearly all the flux will be captured by the fibre.
Each galaxy in the survey, then, acts as a lens magnifying objects
behind, by an amount that depends on the lens mass, the degree of
alignment, and the source and lens redshifts.  The luminous red galaxy
(LRG) subsample (Eisenstein et al.  2001) is particularly useful since
these sources dominate the cross section for strong lensing.
Motivated by these arguments Bolton et al.  (2004, hereafter B04)
searched the spectra of some 51,000 SDSS LRG spectra for anomalous
emission lines superposed on the early-type galaxy spectra.  They
provide a sample of 49 candidate lenses, with source redshifts
confirmed by the detection of at least three lines.  The Einstein ring
J$140228.22+632133.3$ is the first confirmed success from this
programme.

With the same motivation we have searched the spectra of LRGs in the
SDSS DR3 spectroscopic catalogue (Abazajian et al.  2005), to find
galaxy-galaxy lenses.  This paper presents the first two confirmed
lenses from our sample of candidates.  Our survey is similar in many
respects to that of B04, but with some significant differences.  In
\S2 we briefly describe the steps followed to produce a sample of
candidate lenses, highlighting differences compared to B04.  In \S3
and \S4 we describe imaging and spectroscopic observations of two
candidates, confirming that the emission lines are indeed from
multiply-imaged background galaxies.  These observations are adequate
for confirming the nature of the candidates, but insufficient for a
useful analysis of the galaxy mass profiles, which we postpone until
completion of scheduled ACS observations.

When presenting calculations that involve angular diameter distances,
we assume a Friedmann--Robertson--Walker cosmological model defined by
the parameters $\rm \Omega_{M,0} = 0.3$, $\Omega_{\Lambda,0}=0.7$ and
$\rm H_0=70$ km\,s$^{-1}$ Mpc$^{-1}$.

\section{Spectroscopic identification of candidate gravitational lenses}
\label{sec_spec}

We have completed a search for anomalous emission lines in the spectra
of the LRG subsample of the SDSS DR3.  Details of the sample searched,
the line-detection algorithm, and the resulting list of candidates
will be reported elsewhere.  The line-detection algorithm is similar
to that of B04.  Here we outline the main differences.

The first difference between our analysis and that of B04 is the
pre-processing of the SDSS spectra using the algorithm of Wild \&
Hewett (2005), to reduce systematic sky-line residuals.  Large
systematic residuals arise at the wavelengths of the brightest OH sky
lines due to insufficient sampling of the line profiles, the exact
pattern depending on the wavelength solution for each spectrum.  To
deal with this issue B04 modeled the distribution of residuals at each
wavelength, and rescaled the Poisson error distribution, to avoid
detecting sky residuals as false emission lines.  Instead Wild \&
Hewett (2005) have developed a scheme to reduce the systematic 
residuals significantly.  From the large database of spectra from fibres
devoted to sky, the different patterns of residuals can be
characterised and quantified using principal component analysis (PCA),
and accordingly removed from the target spectra.  The noise
characteristics of the resulting spectra approach the Poisson
expectation.  The result of employing the Wild \& Hewett
sky-subtraction procedure is to improve substantially the sensitivity
of line detection at wavelengths $\lambda > 7200$\AA.

To search for anomalous emission lines B04 firstly create a residual
spectrum by subtracting the combination of a best-fit early-type
galaxy spectrum plus a smooth continuum.  The early-type spectrum is
computed as the best-fit redshifted linear combination of PCA
components determined from the sample of restframe LRG spectra.  By
virtue of the well-defined colour selection used in defining the LRG
sample, the continuum emission from any lensed source will be weak
relative to the target LRG.  Therefore the resulting spectra to be
searched contain essentially only emission lines, with any residuals
from absorption lines in the background galaxy well below the noise
level (Figs 1 and 3 in B04).  Instead we have used the simpler scheme
of subtracting a median-filtered version of the spectrum, with a box
length of 41 pixels, corresponding to $\sim60$\AA\, over the
wavelengths of interest.  We have found that the PCA scheme, and
related schemes, produce slightly better results in the region of the
4000\AA \ break, but given the small region affected, and the fact that
we are interested only in detections of relatively high S/N, we prefer
the simplicity of the median-filter scheme.

Emission line identification is then performed on the resulting
spectra using standard matched-filter techniques (e.g.  Hewett et al.
1985).  In contrast to B04 who require at least three emission lines
of concordant redshift, we include candidates with single lines, or
concordant pairs.  BO4's approach essentially ensures that all their
candidates are real, in the sense of detecting a second galaxy at
higher redshift than the LRG \---\ which may or may not be multiply
imaged.  But it limits the highest redshift to $z=0.8$, when [OIII]
$\lambda\lambda$ 4960, 5008 becomes redshifted beyond the red
wavelength limit of the spectra of $9180$\AA.

Targetting single emission line detections, which in most cases would
correspond to the detection of [OII] $\lambda$3728 emission at
redshifts $0.8 < z < 1.4$, might seem an optimistic approach.
However, as B04 discuss, the SDSS spectra have sufficient resolution
that the [OII] $\lambda\lambda$3726.1,3728.8 doublet is partially
resolved, producing a distinctive emission line profile.  For the
identification of genuine lenses, the extension of the search beyond
$z\sim 0.8$ is advantageous because the number of unlensed emission
line galaxies that are bright enough to be detected in the spectra is
very small, i.e.  if a $z > 0.8$ emission line galaxy is detected,
then the probability it is lensed is high\footnote{The faintest
detectable [OII] $\lambda\lambda$3726,3729 emission lines in the LRG
spectra at $z > 0.8$ have fluxes of $9\times10^{-17}{\rm erg}{\rm
s}^{-1}{\rm cm}^{-2}$ while the median emission line flux for the
$\sim 25$ candidate emission lines is $16\times10^{-17}{\rm erg}{\rm
s}^{-1}{\rm cm}^{-2}$.  Typically, as a result both of seeing and the
angular offset between the LRG and an emission line source, the
fraction of the total emission line flux from a star-forming galaxy
contained within the $1\farcs5$ SDSS spectroscopic fibre is $\sim
50\%$, increasing the estimate of the true emission line fluxes by a
factor of $\sim 2$.  The total area of the survey for emission line
objects is equal to the number of galaxies surveyed multiplied by the
effective area of each fibre (for the detection of an emission line
object).  Adopting an effective area of $7\,$square arcseconds per
fibre and a detection efficiency of $50\%$ gives an effective survey
area of $0.02\,$square degrees for sources with intrinsic emission
line fluxes of $18-32 \times10^{-17}{\rm erg}{\rm s}^{-1}{\rm
cm}^{-2}$.

An estimate of the number of unlensed emission line objects predicted
to occur in the LRG spectra can be made using the results of the
Hubble Space Telescope survey for emission line galaxies of Drozdovsky
et al.  (2005).  In an effective area of $0.03\,$square degrees, very
similar to our survey, they identify seven and one object(s) in the
redshift range $0.8 \le z \le 1.3$ with fluxes exceeding $18$ and $32
\times10^{-17}{\rm erg}{\rm s}^{-1}{\rm cm}^{-2}$ respectively.  Thus,
based on the Drozdovsky et al.  findings, the predicted number of
unlensed emission line sources at $z > 0.8$ in our sample is
small.}. In principle, the search for single emission lines also
allows gravitationally-lensed star-forming galaxies with very high
redshifts, $z > 4.5$, to be detected via their Lyman-$\alpha$
$\lambda$1216 emission.  In practice, there appear to be few such
viable candidates in the SDSS DR3 spectroscopic database. Application
of our spectroscopic search technique to 81,610 LRGs selected from the
SDSS DR3 produces $\sim 25$ high S/N candidate $z>0.8$ [OII]
emission-line systems. As of May 2005 we have succeeded in obtaining
follow-up imaging (3 candidates) and spectroscopic (2 candidates)
observations to identify cases of multiple imaging. Here we present
observations of J145957.1-005522.8 and J230946.3-003912.9, confirmed
as gravitational lenses.  For compactness, in the remainder of the
paper we refer to the two targets as J1459 and J2309 respectively.
Both candidates were selected on the basis of the detection of a
single emission line, which, if confirmed as the [OII]
$\lambda\lambda$3726.1,3728.8 doublet, provides redshifts of $z=0.938$
for J1459, and $z=1.005$ for J2309.

\section{Follow--up imaging}
\label{sec_image}

The most direct way to confirm that candidates are lenses would be to
image in the line, by means of observations with an Integral Field
Unit, to confirm the reality of the line, and prove that multiple
images of the source exist.  An alternative approach is to combine
imaging and spectroscopy.  With limited observing time, broadband
imaging on its own may produce ambiguous results, given the scale of
the structures, typical seeing conditions, and the flux ratios between
different copies of the source.  The SDSS images themselves are
inadequate because of the mediocre seeing (1\farcs4 and 1\farcs7 for
J1459 and J2309 respectively) and limited depth.  Longslit
spectroscopy on its own may also be ineffective since there is no
guide to choosing the slit orientation to capture two copies of the
source.  Our strategy has been to obtain images of substantially
better resolution and depth than the original data.  Failing
unambiguous confirmation of lensing from the imaging data, where the
data are suggestive of lensing the images provide a guide to orienting
the slit for spectroscopic confirmation.

\begin{figure*}
\includegraphics[width=110mm]{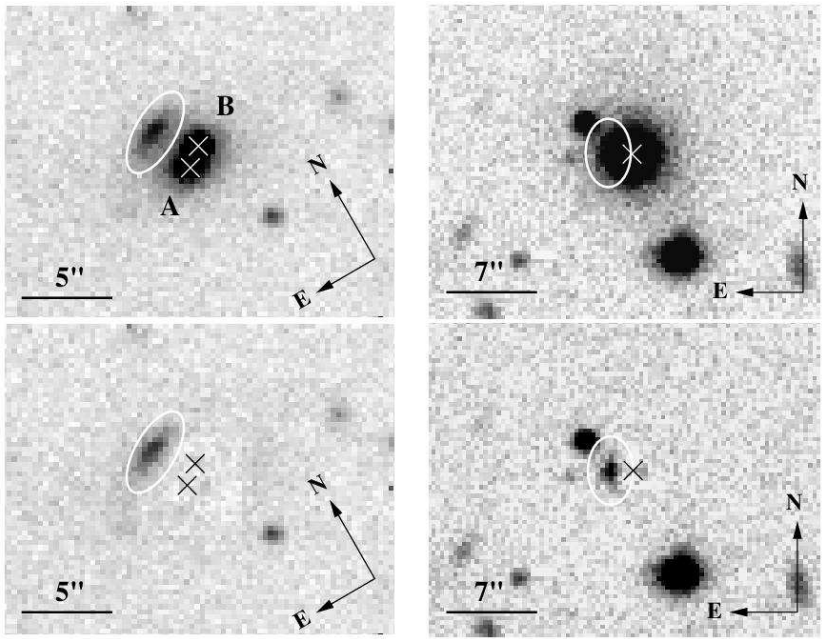}
\caption{Follow--up $r$--band images of the two new lens systems.
Left panels show J1459 and right panels show J2309. In each case the
upper and lower panels show, respectively, the field before and after
subtraction of the images of the lens galaxies. The centroids of the
galaxies subtracted are marked by white crosses (black crosses are
used in the lower panels for better contrast). Note that there are
two lensing galaxies in the field of J1459. The apparent primary
images of the sources are marked by white ellipses. In the case of
J2309 a possible weak counterimage is just visible}
\label{2lens_image}
\end{figure*}

The system J1459 was observed with the MagIC CCD camera at the 6.5m
Clay Telescope on 2003 May 30, in the $r$ band, for 600s, in 0\farcs6
seeing, in clear conditions.  The system J2309 was observed with the
EMMI instrument at the 3.5m ESO NTT on 2004 October 10, in the $r$
band, for 300s, in 1\farcs0 seeing, in clear conditions.  The data
were processed employing standard CCD reduction techniques and the
resulting images are displayed in Fig.  \ref{2lens_image}.  The data
were calibrated using the SDSS photometry of objects detected in the
frames, neglecting any colour term. A zero-point offset applied and
all subsequent photometric measures are quoted in Gunn $r$ using the
AB magnitude scale\footnote{Equivalent Vega $R$ magnitudes can be
obtained by subtracting 0.15 from the AB photometry.}. The $5\sigma$
limiting sky noise computed within a 3\arcsec\ diameter aperture is $r
=24.4$ and 23.9 for the images of J1459 and J2309 respectively.

To subtract the images of the lensing galaxies we used an iterative
procedure as follows (see e.g.  Wayth et al.  2005, for a more
detailed description of the procedure).  We firstly ignored the
contribution to the image of the lensed source, and fit de
Vaucouleurs models, convolved with the point spread function (defined
by stars in the frame), finding the min-$\chi^2$ 6-parameter model
(x, y, orientation, ellipticity, half-light radius $r_{0.5}$, and
surface brightness $\Sigma_{0.5}$ at $r_{0.5}$).  This model was
subtracted, enhancing the image of the source.  Regions covered by the
source were masked, and an improved model computed, etc.  In the case
of J2309 the final fit was unsatisfactory near the centre, so we added
the extra parameter required by the Sersic model, and repeated the
procedure.  In the case of J1459 the image shows two lens galaxies,
separated by $1\farcs2$, so a different procedure was needed.  The two
galaxies are labeled A and B, in Fig.  \ref{2lens_image}.  Rather than
attempt to model both galaxies simultaneously (i.e.  a 12-parameter
fit, for the de Vaucouleurs model), we proceeded by firstly fitting
only the central region of the first galaxy, subtracting the model in
this region, and then fitting the central region of the second galaxy.
This second model was subtracted from the original image, allowing an
improved fit for the first galaxy, over a slightly larger region.
This process was iterated, slowly growing the region of the fits,
while also improving the mask of the image of the source. For both
galaxies, the subtractions are visually satisfactory, and therefore
sufficient for examining the source morphology, but formally, in terms
of $\chi^2$, the fits are unsatisfactory.

Inspection of the image of J1459 reveals an extended arc NNE of the
galaxy pair, offset some $1\farcs8$ from the point midway between the
two galaxies, and aligned tangentially relative to this point.  At the
depth of the image no counterimage is visible, but the tangential
orientation of the arc is strongly suggestive of lensing, so we
flagged the target as high priority for spectroscopy.

The image of J2309 displays a source $1\farcs7$ W of the target.  A
possible faint counterimage is visible $0\farcs7$ E of the target.
The reality of the counterimage is difficult to quantify because of
the possibility of systematic errors in the galaxy subtraction.
Nevertheless the configuration is again strongly suggestive of
lensing, and the target was similarly flagged as high priority for
spectroscopy.

Relevant parameters from the profile fits for the three galaxies are
provided in Table \ref{sersicfits}.  For each galaxy the following
four parameters of the fit are listed: the half-light radius
$r_{0.5}$, the surface brightness at the half-light radius
$\Sigma_{0.5}$, the exponent of the Sersic profile $n$ (i.e. de
Vaucouleurs is $n=4$), and the total magnitude, computed by
integrating the model profile. The uncertainties were computed from
Poisson considerations. Nevertheless changing the size of the fitted
region sometimes produced larger changes in parameters. We defer more
sophisticated modelling of the light profiles until completion of the
ACS observations.  Also listed are the lens redshift, from the SDSS
catalogue, and the source redshift, taken from \S4.  The final row
provides the one dimensional velocity dispersion $\sigma_v$ measured
by SDSS for J2309.  A value is not quoted for J1459, conceivably
because the algorithm was confused in analysing the superposition of
two LRG spectra.

\begin{table}
\caption{Measured properties of the LRGs in each system. No
SDSS velocity dispersion information is available for J1459.}
\label{sersicfits}
\centering{
\begin{tabular}{lccc}\hline
                & \multicolumn{2}{c}{J1459} & J2309 \\ 
                &      A      &      B      &       \\ \hline
  $r_{0.5}$ (arcsec)     &   2.4$\pm0.1$   & 1.6$\pm0.1$ &  2.8$\pm0.1$  \\
 $\Sigma_{0.5}$ ($r$ mag/arcsec$^2$)& 25.7$\pm0.05$ & 25.1$\pm0.05$ & 23.6$\pm0.08$  \\
  $n$           &  (4)        &   (4)       & 4.9$\pm0.13$   \\
  $r_{tot}$(AB)     &   20.5      &    20.7     & 17.8  \\
  $z_l$         & \multicolumn{2}{c}{0.577} & 0.291 \\
  $z_s$         & \multicolumn{2}{c}{0.938} & 1.005 \\
  $\sigma_v$ (kms$^{-1}$) & ... & ... & $197 \pm 18$ \\ \hline
\end{tabular}
}
\end{table}

Magnitudes for the putative primary source image in each system were
computed within 3\arcsec\ (J1459) and a 2\arcsec (J2309) diameter
circular apertures centred on the source in the LRG--subtracted image
\---\ generating $r$--band magnitudes of $22.5$ for J1459 and $22.3$
for J2309.  The presence of a nearby, unrelated, source in J2309
necessitating the use of the smaller aperture.

\section{Long--slit spectroscopy}
\label{sec_lss}

To confirm that the source in each system is multiply imaged, we
obtained spectra with the slit oriented from the centre of the lens to
the candidate primary image of the source.  For J1459 the lens centre
was defined as the flux-weighted centroid of the two galaxies.  The
goal of the observations was to detect the emission line on either
side of the LRG.

The systems J1459 and J2309 were observed with the RILD configuration
of the EMMI instrument at the ESO NTT on 2004 June 6, and October 10
respectively, in clear conditions.  The 600 lines mm$^{-1}$ grism with
a $1\farcs0$ slit provided spectral coverage from 5800 to 8500\AA\, at
a resolving power $\sim 1500$.  The spatial scale was 0\farcs33
pix$^{-1}$.  The slit orientations, the exposure times, and the
average seeing during the observations, are provided in Table
\ref{tab_lens_values}.  Conventional procedures were followed for bias
subtraction and flatfielding.  Then the frames for each target were
averaged, employing a sigma--clipping algorithm in order to remove
cosmic ray events.

\begin{figure}
\includegraphics[width=84mm]{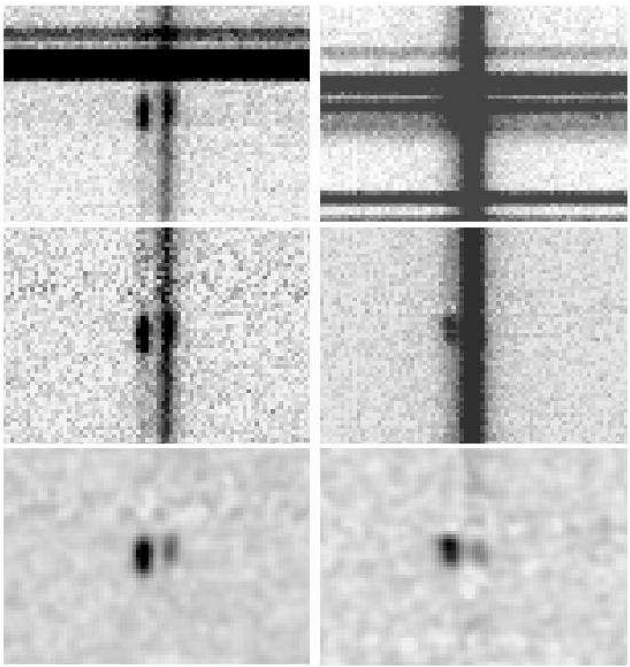}
\caption{Spectroscopy of J1459 (left column) and J2309 (right
 column). The three rows correspond to different stages of the
 analysis. The spectral axis is oriented vertically and the spatial
 axis is oriented horizontally: the displayed dimension of each panel
 is approximately 24\arcsec $\times$ 80\AA. {\em Top row: } Before sky
 subtraction, {\em middle row: } after sky subtraction, {\em bottom
 row: } after continuum subtraction of sources, and convolution with a
 Gaussian of $\sigma=1$ pixel.}
\label{lens1_spec_panel}
\end{figure}

Small sections of each of the final frames, at various stages of
subsequent processing, are shown in Fig.  \ref{lens1_spec_panel}.  In
this figure the left column of panels shows J1459, and the right
column shows J2309.  The top row shows the spectra before sky
subtraction, and the middle row shows the spectra after sky
subtraction.  The bottom row is explained below.  Referring to the
middle row, the dark vertical line in each panel is the spectrum of
the LRG(s).  For both targets an emission line, spatially offset from
the LRG, is clearly visible, confirming the original detection of an
emission line in the SDSS spectrum.  In the case of J2309 the line is
visibly double peaked, and the wavelength separation confirms the
identification as the [OII] $\lambda\lambda$3726.1,3728.8 doublet at
$z=1.005$.  In the case of J1459 the line is spectrally resolved, and,
although not clearly double peaked, the profile is suggestive of the
[OII] doublet at $z=0.938$.  As described below, detection of the
corresponding H$\gamma$ line confirms this identification.  The source
redshifts have been entered into Table \ref{sersicfits}.

To subtract the spectrum of the lens galaxy in each system, in order
to search for a counterimage of the emission line, we fit a cubic
spline, with a small number of pieces, up each column.  This is
satisfactory because there are no strong absorption lines in the LRGs
at the wavelengths of the emission lines.  The resulting spectra,
smoothed by convolution with a Gaussian, are reproduced in the bottom
row of Fig.  \ref{lens1_spec_panel}.  An emission line on the opposite
side of the LRG, and at the same wavelength as the line from the
primary image, and therefore corresponding to a secondary image of the
source, is visible in both panels.  The significance of the detection
is $9.5\sigma$ for J1459, and $4.6\sigma$ for J2309, established
through measurement of the unsmoothed frames.  These detections,
corroborated by the measured source redshifts $z_s>z_l$, confirm that
we have identified galaxies that are multiply imaged, and that both
systems are gravitational lenses.  The spatial separations of the two
images of the [OII] $\lambda$3728 lines of each source, relative to
the LRG, as well as the flux ratios of the primary and secondary
images, were measured in the unsmoothed continuum-subtracted frames.
The results are summarised in Table \ref{tab_lens_values}.  

For each confirmed lens system, the location of the lensed
emission-line images about the LRG deflector in the two-dimensional
spectral frame imply that the total magnification is low: inputting
the deflector--image separations, source and deflector redshift for
each system into an assumed singular isothermal spherical (SIS) mass
model generates total magnification factors 2.8 (J1459) and 4.0
(J2309) for each lens system. Low magnification values for the
secondary images (0.4 and 1.0 for J1459 and J2309 respectively)
provide some explanation for the lack of detection in the $r$-band
image, Fig.  \ref{2lens_image}. The predicted flux ration $A/B=6$ for
J1459 compares to the value of 2.6 measured from the long--slit
spectrum. This discrepancy is not thought to be serious as the primary
lens image is extended and slit losses will reduce the measured flux
ratio. We note however, that computation of lens properties based upon
the assumption of a simple mass model must be treated with some
caution: the SIS mass model predicts an Einstein ring diameter of
$1\farcs5 \pm 0.25$ for the geometry of the J2309 system. The observed
image separation in the spectrum, $2\farcs34 \pm 0.46$ is consistent
with this estimate, but this comparison is not particularly
illuminating given the large uncertainties.

\begin{table}
\caption{Details of the spectroscopic observations of each system, and
the measured parameters for the detected [OII] $\lambda$3728 lines.}
\label{tab_lens_values}
\begin{tabular}{lcc}
\hline
& J1459 & J2309 \\
\hline
Slit position angle ($^{\circ}$EofN) & 26.1 & 90.0 \\
Exposure time (s)    & $3\times1800$  & $4\times 1800$ \\
Average seeing (\arcsec) & 0.8 & 1.1 \\
Image separation (\arcsec) & $2.11 \pm 0.26$ & $2.34 \pm 0.46$ \\
$|\rm Galaxy - A|$ (\arcsec) & $1.81 \pm 0.18$ & $1.75 \pm 0.15$ \\
$|\rm Galaxy - B|$ (\arcsec) & $0.30 \pm 0.30$ & $0.59 \pm 0.45$ \\
Flux ratio A/B & $2.58 \pm 0.34$ & $3.16 \pm 0.53$ \\
\hline
\end{tabular}
\end{table}

The spectra of the LRG and both the primary and secondary images of
the emission line galaxy overlap.  To separate and extract the spectra
of the lens and source, for each system, we used the following
simplified scheme.  We firstly fixed the relative locations of the
lens and the two images of the source, in each system, at the values
listed in Table \ref{tab_lens_values}.  We then defined the normalised
cross-sectional profile for the source and lens separately.  The
source cross-section was defined by fitting a Gaussian to the spatial
profile of the primary image of the [OII] $\lambda$3728 line, and
using the same profile for the secondary image of the source, with the
flux ratio fixed at the value measured for the emission line.  The
profile of the lens was defined by fitting a Gaussian to the peak of
the cross-section, over a wavelength range separate from the [OII]
$\lambda$3728 line.  The problem of spectral extraction then reduces
to the problem of a min-$\chi^2$ fit at each wavelength, with three
free parameters: the position of the lens, and the normalisations of
the lens and source profiles (which provide the fluxes).
Corresponding variance spectra were computed from Poisson
considerations.

The above procedure represents a relatively robust approach to the
problem of spectral extraction for blended, moderate S/N data.
Inspection of the residual frame generated by subtracting the
min-$\chi^2$ fit reveals coherent residual light at the level
typically equal to or lower than the $1\sigma$ errors.  These features
arise from the fact that the adopted Gaussian
spatial profile does not provide an exact match to the LRG in each
system (the brightest spectral component).  However, it is not
anticipated that further detailed modelling of the LRG light
distribution will result in substantive changes to the resulting
spectra.

\begin{figure}
\includegraphics[width=64mm,angle=270.0]{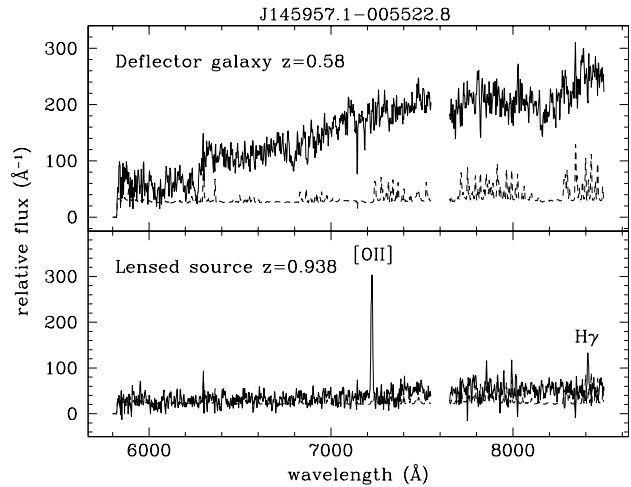}
\caption{One dimensional extracted spectra for the lens system
J1459. The top panel displays the extracted spectrum of the LRG and
the lower panel displays the extracted spectrum of the source. In each
panel the data (solid line) and $1\sigma$ noise (dashed line) spectra
are displayed. Data at wavelengths associated with terrestrial
atmospheric absorption ($7550<\lambda<7650$) are not plotted. In the
lower panel, the locations of the detected [OII] $\lambda$3728 and
H$\gamma$ lines are marked.}
\label{1459_2panel}
\end{figure}

\begin{figure}
\includegraphics[width=64mm,angle=270.0]{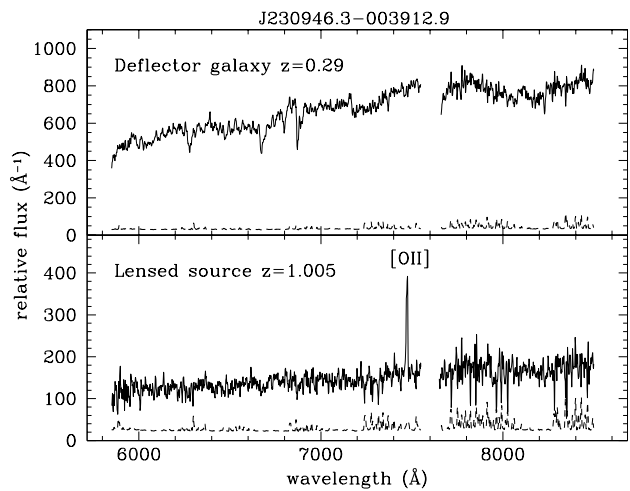}
\caption{One dimensional extracted spectra for the lens system
J2309. The top panel displays the extracted spectrum of the LRG and
the lower panel displays the extracted spectrum of the source. In each
panel the data (solid line) and $1\sigma$ noise (dashed line) spectra
are displayed. Data at wavelengths associated with terrestrial
atmospheric absorption ($7550<\lambda<7650$) are not plotted. In the
lower panel, the location of the detected [OII] $\lambda$3728 line
is marked.}
\label{2309_2panel}
\end{figure}

The extracted flux-calibrated spectra, on a linear wavelength scale,
are displayed in Figs \ref{1459_2panel} and \ref{2309_2panel}.  We
detect the corresponding H$\gamma$ emission line in the spectrum of
J1459, at $4\sigma$ significance, which confirms our original
identification of the line detected in the SDSS spectrum as [OII]
$\lambda$3728 at $z=0.938$. The H$\gamma$ emission line for J2309 is
not detected in the SDSS DR3 survey spectrum above the nominal flux
limit of $5\times 10^{-17}$ ergs s$^{-1}$ cm$^{-2}$ and it lies
outwith the accessible wavelength interval for the NTT/EMMI follow-up
spectrum. The unreddened [OII] $\lambda$3728 emission-line fluxes
observed in the SDSS DR3 spectra are $13.9 \times 10^{-17}$ ergs
s$^{-1}$ cm$^{-2}$ (J1459) and $9.4 \times 10^{-17}$ ergs s$^{-1}$
cm$^{-2}$ (J2309). These fluxes are lower limits due to unknown signal
losses arising from the limited spectroscopic fibre field of
view. Employing the total magnification factors computed for each
system, the corresponding unlensed [OII] $\lambda$3728 emission
luminosity for each system is $2.1 \times 10^{41}$ ergs s$^{-1}$
(J1459) and $1.3 \times 10^{41}$ ergs s$^{-1}$ (J2309). Employing the
conversion between [OII] $\lambda$3728 emission luminosity and star
formation rate (SFR) presented by Kewley et al. (2004), the intrinsic
SFR detected in each lensed source is approximately 1.4 (J1459) and
0.8 (J2309) M$_\odot$ yr$^{-1}$.


In summary, imaging and spectroscopic observations of two candidate
gravitational lenses, selected through the detection of anomalous
emission lines in the spectra of SDSS LRGs, have resulted in the
confirmation of two new galaxy--galaxy lenses. The new lenses are
suitably bright for HST imaging. Only three similarly bright
galaxy--galaxy lenses were previously known. ACS observations can
provide accurate measurement of the dark matter density profile in the
lens galaxy in such systems, free of dynamical assumptions. Many more
similarly--good candidate systems exist in the SDSS database, providing
the basis for detailed statistical studies of galaxy dark matter
profiles.

\section*{acknowledgements}

Funding for the Sloan Digital Sky Survey (SDSS) has been provided by
the Alfred P. Sloan Foundation, the Participating Institutions, the
National Aeronautics and Space Administration, the National Science
Foundation, the U.S. Department of Energy, the Japanese
Monbukagakusho, and the Max Planck Society. The SDSS Web site is
http://www.sdss.org/.

The SDSS is managed by the Astrophysical Research Consortium (ARC) for
the Participating Institutions. The Participating Institutions are The
University of Chicago, Fermilab, the Institute for Advanced Study, the
Japan Participation Group, The Johns Hopkins University, Los Alamos
National Laboratory, the Max-Planck-Institute for Astronomy (MPIA),
the Max-Planck-Institute for Astrophysics (MPA), New Mexico State
University, University of Pittsburgh, Princeton University, the United
States Naval Observatory, and the University of Washington.

The authors are additionally grateful to Thodoris Nakos for having
performed the June 2004 NTT observations. We further thank the
anonymous referee for a constructive report.

\bsp

\label{lastpage}

\end{document}